**Climate Change Research in View of Bibliometrics**

Robin Haunschild*[+], Lutz Bornmann**, Werner Marx*

* Max Planck Institute for Solid State Research
Heisenbergstraße 1, 70569 Stuttgart, Germany
E-mail: r.haunschild@fkf.mpg.de, w.marx@fkf.mpg.de

** Division for Science and Innovation Studies
Administrative Headquarters of the Max Planck Society
Hofgartenstr. 8,
80539 Munich, Germany.
E-mail: bornmann@gv.mpg.de

[+] Corresponding author

**Abstract**

This bibliometric study of a large publication set dealing with research on climate change aims at mapping the relevant literature from a bibliometric perspective and presents a multitude of quantitative data: (1) The growth of the overall publication output as well as (2) of some major subfields, (3) the contributing journals and countries as well as their citation impact, and (4) a title word analysis aiming to illustrate the time evolution and relative importance of specific research topics. The study is based on 222,060 papers (articles and reviews only) published between 1980 and 2014. The total number of papers shows a strong increase with a doubling every 5-6 years. Continental biomass related research is the major subfield, closely followed by climate modeling. Research dealing with adaptation, mitigation, risks, and vulnerability of global warming is comparatively small, but their share of papers increased exponentially since 2005. Research on vulnerability and on adaptation published the largest proportion of very important papers (in terms of citation impact). Climate change research has become an issue also for disciplines beyond the natural sciences. The categories *Engineering* and *Social Sciences* show the strongest field-specific relative increase. The *Journal of Geophysical Research*, the *Journal of Climate*, the *Geophysical Research Letters*, and *Climatic Change* appear at the top positions in terms of the total number of papers published. Research on climate change is quantitatively dominated by the USA, followed by the UK, Germany, and Canada. The citation-based indicators exhibit consistently that the UK has produced the largest proportion of high impact papers compared to the other countries (having published more than 10,000 papers). Also, Switzerland, Denmark and also The Netherlands (with a publication output between around 3,000 and 6,000 papers) perform top - the impact of their contributions is on a high level. The title word analysis shows that the term *climate change* comes forward with time. Furthermore, the term *impact* arises and points to research dealing with the various effects of climate change. The discussion of the question of human induced climate change towards a clear fact (for the majority of the scientific community) stimulated research on future pathways for adaptation and mitigation. Finally, the term *model* and related terms prominently appear independent of time, indicating the high relevance of climate modeling.

**1. Introduction**

Climate change is a change in the statistical distribution of weather patterns during an extended period of time (from decades to millions of years). Meanwhile, *climate change* and *global warming* are terms for the observed century-scale rise in the average temperature of the earth's surface. From the perspective of large time periods, climate change is caused by a multitude of factors like variations in solar radiation (changing parameters of the earth's orbit, variations of the solar activity observed via sunspot number), drifting continents (see plate tectonics), volcanic eruptions (producing large amounts of sulfate-based aerosols), and possibly others. During the last decades, human activities (in particular burning of fossil fuel and pollution as the main consequences of the growth of population and industrialization) have been identified as significant causes of recent climate change, often referred to as *global warming*. The Intergovernmental Panel on Climate Change (IPCC) reports in its foreword that "the IPCC is now 95 percent certain that humans are the main cause of current global warming (IPCC Synthesis Report 2014, p. v; see also Anderegg et al., 2010). The report states in its *Summary for Policymakers* that "human influence on the climate system is clear, and recent anthropogenic emissions of greenhouse gases are the highest in history. Recent climate changes have had widespread impacts on human and natural systems" (IPCC Synthesis Report 2014, Summary for Policymakers, p. 2).

Climate change has gained strongly increasing attention in the natural sciences and more recently also in the social and political sciences. Scientists actively work to understand the past climate and to predict the future climate by using observations and theoretical models. Various subfields from physics, chemistry, meteorology, and geosciences (atmospheric chemistry and physics, geochemistry and geophysics, oceanography, paleoclimatology etc.) are interlinked. Climate change has also become a major political, economic, and environmental issue during the last decade and a central theme in many political and public debates. How to address, mitigate and adapt to climate change has become a hot issue. The scientific community has contributed extensively to these debates with various data, discussions, and projections on the future climate as well as on the effects and risks of the expected climatic change.

The large and strongly growing amount of climate change research literature has brought about that scientists working within this research field experience increasingly problems to maintain a view over their discipline. Modern information systems could possibly offer databases and analysis tools providing a better overview on the entire research field. However, due to lack of access and experience concerning suitable databases and analysis tools, many experts do not take advantage of them. But meanwhile, a series of bibliometrics analyses have been published by scientometricians, stimulated by the growing scientific, political, and public attention of research on climate change. These publications appeared both in subject specific journals in the field of climate change research as well as in bibliometrics journals:

Based on a sample of 113,468 publications on environmental assessment from the past 20 years, Li and Zhao (2015) "used a bibliometric analysis to study the literature in terms of trends of growth, subject categories and journals, international collaboration, geographic distribution of publications, and scientific research issues" (p. 158). The h index was used to evaluate global environmental assessment research quality among countries. According to this study, the USA and UK have the highest h index among the contributing countries. Stanhill (2001) discussed the growth of climate change science and found a doubling of the related publications every 11 years. Li et al. (2011) analyzed the scientific output of climate change research since 1992 "to assess the characteristics of research patterns, tendencies, and methods in the papers… It was concluded that the items 'temperature', 'environment', 'precipitation', 'greenhouse gas', 'risk', and 'biodiversity' will be the foci of climate change research in the 21st century, while 'model', 'monitoring', and 'remote sensing' will continue to be the leading research methods" (p. 13). Based on co-citation analysis, Schwechheimer and Winterhager (1999) identified highly dynamic, rapidly developing research fronts of climate research. ENSO (El Nino Southern Oscillation) irregularity, vegetation & ice-age climate, and climate-change & health were found as the research fields with the highest immediacy values.

Other bibliometric studies deal with more specific topics within the field of climate research: Ji et al. (2014) analyzed research on Antarctica, Wang et al. (2014) discussed the vulnerability of climate change, and Pasgaard and Strange (2013) presented a quantitative analysis of around 15,000 scientific publications from the time period 1999-2010, discussing the distribution of climate change research throughout the contributing countries and the potential causes of this distribution. Vasileiadou (2011) have explored the impact of the IPCC Assessment Reports on science. Bjurström and Polk (2011a, 2011b) analyzed the interdisciplinarity of climate change research based on the referenced journals in the IPCC Third Assessment Report (2001) via co-citation analysis. Hellsten and Leydesdorff (2015) analyzed the development of the knowledge base and programmatic focus of the journal *Climatic Change*. Most interesting and in contrast to substantial public doubt are the findings of Anderegg et al. (2010). They conclusively revealed the striking agreement among climate research scientists on the anthropogenic cause of climate

5change based on the publications of 1,372 top climate experts. Janko et al. (2014) analyzed the controversies about climate change through comparison of references in and citations of contrarian reports.

Most of the studies in the past focused on specific topics within climate change research and do not present an analysis of the complete research field. The few comprehensive studies mentioned above used search queries for the literature search which are more or less inappropriate: They are mostly based on queries for literature retrieval, using somewhat arbitrary items for selecting subfields. The corresponding publication sets are therefore limited with regard to completeness: Stanhill (2001) has analyzed the growth of climate change relevant literature using exclusively the abstract journal of the *American Meteorological Society* as publication set. Li et al. (2011) analyzed trends in research on global climate change: "'Climate change', 'climate changes', 'climatic change', and 'climatic changes' were used as the keyword to search titles, abstracts, and keywords from 1992 to 2009" (p. 14). Pasgaard and Strange (2013) used "the search phrases climat* AND change* and global warming (with asterisk wildcards)" (p. 1685). Only Wang et al. (2014) applied a more sophisticated method (see below) but his analysis deals exclusively with the climate change vulnerability.

The analysis presented here extends to the time period of the publications relevant for climate change research from 1980 (the time when climate change emerged as a new research field) to the present (end of 2014). We developed a sophisticated search query to cover the relevant literature as completely as possible and to exclude (climate) research not relevant for the global warming issue. Based on a carefully selected publication set of 222,060 papers (including 10,932,050 references cited therein), we firstly analyzed the growth of the overall publication output and of major subfields between 1980 and 2014. Secondly, we examined the topical shifting of the climate change relevant research by title word analysis. Finally, we identified the most contributing journals and countries and their overall citation impact. The previous papers either did not consult any citation based impact data, or they present citation counts which are not normalized with regard to the publication year and the specific research field of the cited publication (e.g. Li and Zhao, 2015).

## 2. Methods

### 2.1 Search for the literature and description of the dataset

It is not an easy task to select all papers[1] related to a specific research field or research topic using literature databases as information source. Completeness or recall (*all* relevant papers) and high precision (*only* relevant papers) are inversely related and mutually exclusive (Buckland and Gey, 1994). This basic connection between completeness and precision precludes a much "cleaner" publication set (i.e. more hits and concurrently less non-relevant papers). In particular, a broad research field like climate change research is not clearly defined and there is no sufficient categorization by keywords, index terms or thesauri. Since the beginning of climate change research, a lot of (neighboring) disciplines tend to relate their research topics on climate research but their papers often deal primarily with other topics.

Wang et al. (2014), who gave an overview of climate change vulnerability, applied a more sophisticated method which they called a four step backward searching. This strategy comprises a preliminary search for key papers and a renewed search based on the synonyms revealed by the keyword analysis of the key papers. This kind of strategy has been called "interactive query formulation" and was discussed extensively by Wacholder (2011): "Iterative query reformulation involves creation of a new variant (reformulation) of a previous query … In the flow-of-information model, query reformulation is treated as a subprocess of the broader QF process" (p. 161). In the present analysis we applied a similar approach for the data retrieval. We have used the Web of Science (WoS) custom data of the database producer Thomson Reuters (Philadelphia, USA) derived from the Science Citation Index Expanded (SCI-E), Social Sciences Citation Index (SSCI), and Arts and Humanities Citation Index (AHCI), allowing more advanced retrieval options than the online version of the WoS.

Step 1: We searched for the term "*climat* chang*" (to include in particular the following phrases: climate/climatic change/changes/changing) within the titles only to establish a publication set of key papers (n = 29,396). Out of this publication set the keywords have been selected and ranked according to their frequency of occurrence. Based on the most frequently appearing more complex keywords within the selected set of key papers we looked for climate change synonyms and established the following list of search terms (with asterisk wildcards for truncation to cover in particular the terms given in parentheses):

- *climat* chang* (climate/climatic change/changes/changing)
- *climat* warming* (climate/climatic)

---

[1] To avoid confusion with the document type "article," the term "paper" rather than "article" is used throughout this manuscript for any kind of journal-based publication.

- *global temperature* (temperature/temperatures)
- *global warming*
- *greenhouse gas* (gas/gases)
- *greenhouse effect* (effect/effects)
- *greenhouse warming*

During selection of these search terms we had to distinguish between "classical" climate research (not referring to global warming) and climate change research (although no clear differentiation is possible). Classical climate research deals for example with the modification of landscapes through glacial periods (ice ages) or with basic topics in meteorology. The keywords have been selected here against the climate change research background. For example, terms like "climate variability" are not included in our search term list, because they appear also in biological and medical studies far from research on climate change. We may assume, however, that papers on climate variability, which are actually relevant for climate change research, are covered by the other search terms.

Step 2: We searched the more complex search terms derived from step 1 listed above and in addition the short term "*climat*" each within the titles only. The left truncation in addition to the right truncation of the term "climat" was used to include also terms such as "pal(a)eoclimate". We found only one single term covered by this left and right truncation which is not closely related to climate change research: "acclimation" or "acclimatization", respectively. Papers with "*acclimat*" in the title were removed unless they contain the term "climat*" (with right truncation but without left truncation) in the title.

Searching titles for the term "*climat* retrieves a certain amount of papers with limited relevance for the climate change topic, but also many papers which mention unforeseeable terms around climate change (e.g. climate cycle, climate model, climate policy, and past climate). Considering the strongly increasing number of publications dealing with research on climate change (see below), we may assume (at least for the more recent publications) a high probability that a paper is related to climate change research, if the term "*climat*" does appear in the title.

Step 3: We searched the more complex terms from step 1 within the abstracts only. Although searching in abstracts only is not possible in the WoS, it is possible in our in-house database derived from the WoS data. Abstract searching based on short terms like "climat*" results in too many papers which are not closely related to climate change research. Note that abstracts are included in WoS since 1991 only.

Step 4: In addition to the title word searching of step 2, we also executed a keyword-based search with the same terms as for the titles. Again, papers with the keywords "acclimation" or "acclimatization" were removed from the publication set unless the keyword section also contained "climat*" (with right but without left truncation).

Step 5: The results from steps 2-4 were combined with a logical OR and refined to articles and reviews as document types (i.e. only substantial contributions to the field are considered). The search as described above was restricted to the time period from 1980 to 2014 and eventually resulted in a publication set of 222,060 papers (articles and reviews only). This is definitely not the complete publication set covering any research paper relevant for the climate change research topic. However, we may assume that we have included by far most of the relevant papers, in particular the key papers dealing with research on climate change. Previous studies dealing with overall climate change research and extending to more recent publication years are based on substantially smaller publication sets (e.g. Li et al. (2011): around 30,000 papers).

For the growth analysis of some major research fields, the publication set resulting from step 5 has been combined with a logical AND using the additional search terms or phrases, respectively, shown in Table 1.

Table 1: Definition of subtopics by search terms which are connected with a logical AND with the climate change literature searched before.

| **Subtopic** | **Search terms** |
| --- | --- |
| Atmosphere | *atmospher* / *aerosol* / *cloud* / *wind* / *storm* |
| Ice & Snow: | *ice* / *glacier* / *snow* / *frost* |
| Oceanic Water | *ocean* / *sea* / *marin* |
| Continental Water | *lake* / *river* / *flood* / *precipitat* / *rainfall* |
| Ocean Currents | *el nino* / *elnino* / *southern oscillation* / *enso* / *Walker circulation* / *north atlantic oscillation* / *nao* |
| Biomass | *biomass* / *agricultur* / *food* / *soil* / *forest* / *plant* / *species* / *vegetat* |
| Climate Model | *model* / *calculat* / *simulat* / *predict* |
| Adaptation | *adapt* / *mitigat* |
| Impacts | *effect* / *impact* |
| Vulnerability | *risk* / *vulnerab * |

These terms were searched in the title and keyword fields. The left and right truncations of these search terms are expected to yield no unwanted hits because each of them is combined via a logical AND with our primary climate change publication set.

**2.2 Statistics for the data analysis**

2.2.1 Normalization of citation impact

The publication set has been analyzed with regard to the most perceived sub-fields, journals and countries by analyzing the overall citation impact of the publications. Thus, we used citation counts to measure the impact publications from certain sub-fields, journals, and countries have on science. Since the total citation impact of the publications is used in this study, not only the impact of the publications on the climate change research is measured, but also the impact on science in general. However, most of the citation impact will fall on the climate change research itself and can thus be interpreted accordingly.

Pure citation counts of papers are not meaningful, because they depend not only on the importance of research (for the research of other researchers than the authors), but also on the subject category and the publication year of the papers (Marx and Bornmann, 2015). For example, one can expect more citations on average for papers in biology than for papers in the social sciences (using citation data from WoS). Therefore, we present citation impact scores in this study which are normalized concerning the particular publication year and WoS subject category (Bornmann and Marx, 2015). The normalization is done in our in-house database as follows: The proportion of papers of a given publication set A (e.g. a journal or a country) which belong to the most frequently cited papers in the corresponding WoS subject categories and publication years has emerged as the most robust normalized impact score (Wilsdon et al., 2015; Wouters et al., 2015). In order to ascertain the proportion of papers, for every paper in A the reference set with comparable papers is composed. The comparable papers consist of papers belonging to the same subject category and publication year as the paper from A. The papers in the reference sets are sorted in descending order by their citation counts and the most frequently cited papers are identified. Then, either a paper from A belongs to the most frequently cited papers in the corresponding reference set ($P_{Ai}=1$) or not ($P_{Ai}=0$). Only some papers from A are fractionally assigned to the most frequently cited papers, if their citation counts position them at the threshold which is used to separate the most frequently cited papers from the rest in the reference set (Waltman & Schreiber, 2013). Based on this value for every paper i in A (0, …, 1), the proportion of papers in A is calculated which belongs to the most frequently cited papers in their set of comparable papers. In the case that a paper belongs to multiple subject categories, impact values are calculated for such papers in each subject category and average values are used for impact values on a paper basis.

The indicator $PP_{top\ 50\%}$ is the proportion of papers in A which belongs to the 50% most frequently cited papers. Thus, this is the proportion of papers which is cited equal to or more frequently than "an average paper" in the corresponding reference sets. $PP_{top\ 50\%}$ is the basic citation impact indicator in bibliometrics which indicates above average perceptions of literature with $PP_{top\ 50\%}>50$ and below average perceptions with $PP_{top\ 50\%}<50$ (50 is the expected value indicating an average impact). Besides $PP_{top\ 50\%}$, two further indicators are frequently used in bibliometric studies (e.g. in the Leiden Ranking), which focus on the excellence level: $PP_{top\ 10\%}$ and $PP_{top\ 1\%}$, respectively, specifies the papers which belong to the 10% and 1%, respectively, most frequently cited papers. The expected values for an average paper set are 10 for $PP_{top\ 10\%}$ and 1 for $PP_{top\ 1\%}$. These indicators show, e.g., whether a given country B was able to publish more papers in the excellence area than a given country C. With more papers in this area, country B would have contributed more important papers to the climate change research than country C.

As citations accumulate rather slowly over time, we restrict the citation impact analyses to the time period 1980-2012 whenever impact indicators are performed. Citations were gathered until May 15$^{th}$ 2015 which allows for a citation window of at least three years.

2.2.2 Mapping of research topics

Bibliometrics aims to quantify the outcome and interconnection of scientific activity. The number of publications is the most popular measure of output, while the number of citations is the most popular indicator of impact, which is one (measurable) aspect of quality. Text searching (data mining) is another tool which may be used to quantify content. A simple method for revealing the hot topics of a research field is based on an analysis of the title words (or alternatively: the keywords) of the literature published so far. In our study we used the VOSviewer software package (Van Eck and Waltman, 2010) for mapping the title words of the climate research literature of our publication set (see www.vosviewer.com)

The maps created by VOSviewer and used in this study are based on bibliographic coupling as a technique to position nodes (in our case: the corresponding title words). The distance between the nodes is proportional to the similarity (relatedness) with regard to the cited references. Hence, title words of papers that cite similar literature are found closer to each other. The automatically performed arrangement of the nodes is highly sensitive and might change significantly if rather few papers are added or removed. However, the size of the nodes (the larger the number of papers with a specific title word, the larger the node) and the distances between each other are hardly affected. The nodes on a map are also assigned by VOSviewer to clusters (they are highlighted in different colors). These clusters identify closely related nodes, where each node is assigned to only one cluster (van Eck & Waltman, 2014). VOSviewer uses a modularity-based clustering technique, which is closely related to the multidimensional scaling technique (Waltman, van Eck, & Noyons, 2010) and is based on the smart local moving algorithm (Waltman & Eck, 2013).

## 3. Results

### 3.1 Overall growth and growth in terms of disciplines and subfields

As a first step to provide an overview of the development of the entire research dealing with climate change, the time evolution of the publication productivity (output) of this research field (measured as number of papers published per year) has been analyzed. FIG 1 shows the annual number of papers published within the time period 1980-2014 and covered by the WoS database.

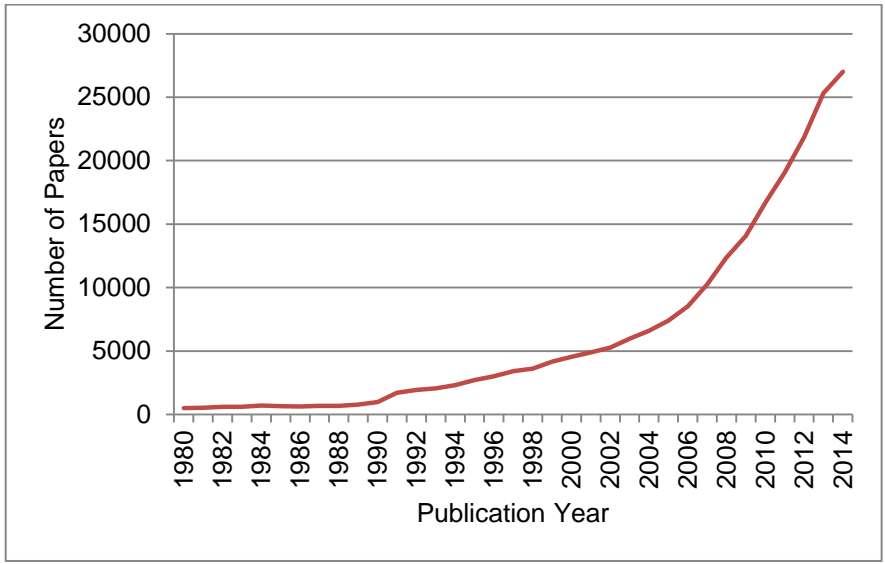

FIG 1: Time evolution of the overall number of climate change relevant papers (articles and reviews). The distinct step from 1990 to 1991 does not signal a sudden increase in productivity but solely the fact that the WoS does not include abstracts before 1991. As a consequence, searches using the WoS field tag "topic" in combination with publication years prior to 1991 yield significantly lower hit numbers (frequently one order of magnitude).

According to FIG 1, the total number of papers dealing with climate change shows a strong increase: Within the time period 1991 to 2010, the number of climate change papers published per year increased by a factor of ten, whereas in the same time period the overall number of papers covered by the WoS databases increased "only" by a factor of around two. The data row of FIG 1 exhibits a doubling of the climate change papers every 5-6 years. The exponential growth of climate change literature is possibly induced by the increasing influence of the IPCC Assessment Reports, which eventually made climate change research a hot topic. These reports revealed the strong need of further research for a better understanding of the earth's climate system and for improved predictions of the future climate. Furthermore, the effects, impacts and risks of climate change became more and more concrete. The discussion of the question of human induced climate change towards a clear fact (at least for the majority of the scientific community, see Anderegg, 2010) stimulated research on future pathways for adaptation and mitigation.

The literature growth is roughly in accordance with the results of Grieneisen and Zhang (2011), who report that the number of publications on climate change and global warming has doubled with a rate of approximately every 4 years. As mentioned above, Stanhill (2001) found a doubling rate of 11 years for the time period 1951-1997. To put this into perspective, we compare our results with the growth rate of the overall science: According to Bornmann and Mutz (2015) the total volume of publications covered by the WoS between 1980 and 2012 doubled approximately every 24 years. Hence, the growth rate of climate change related publications is extraordinarily high. The bend down around 2012 is presumably caused by still incomplete coverage of the recent publication years through WoS and is no sign of decline.

The results of the analysis of the climate change related papers with regard to their disciplines of origin are shown in FIGs 2 and 3. The figures are based on the main OECD categories assigned. Compared to the WoS research areas and subject categories, the main OECD categories are broader grained and therefore better suitable for an overview.

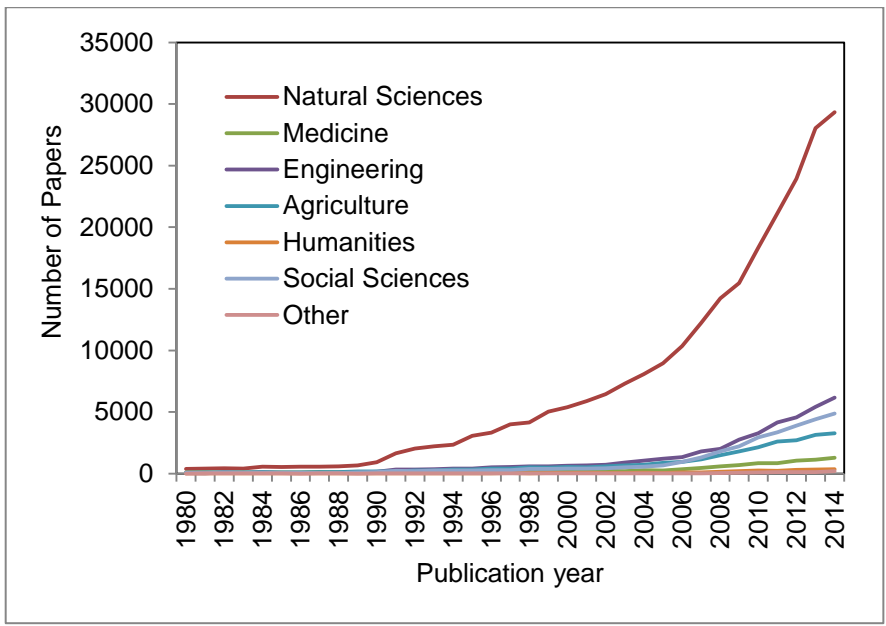

FIG 2: Time evolution of the field-specific climate change related papers published since 1980, based on the main OECD categories assigned. Papers which are assigned to more than one main OECD category are multiply counted.

As expected, climate change research is dominated by the natural sciences. Further analyses of our data with regard to the specific research areas show that the earth sciences (meteorology and atmospheric sciences), the biological, the agricultural sciences, and the environmental sciences are predominant.

FIG 3 shows the relative increase of papers since 1980 assigned to the main OECD categories. The paper share is presented in percent increase based on the numbers from 1980 (thus, the number of papers published in the year 1980 in each case equals 100%).

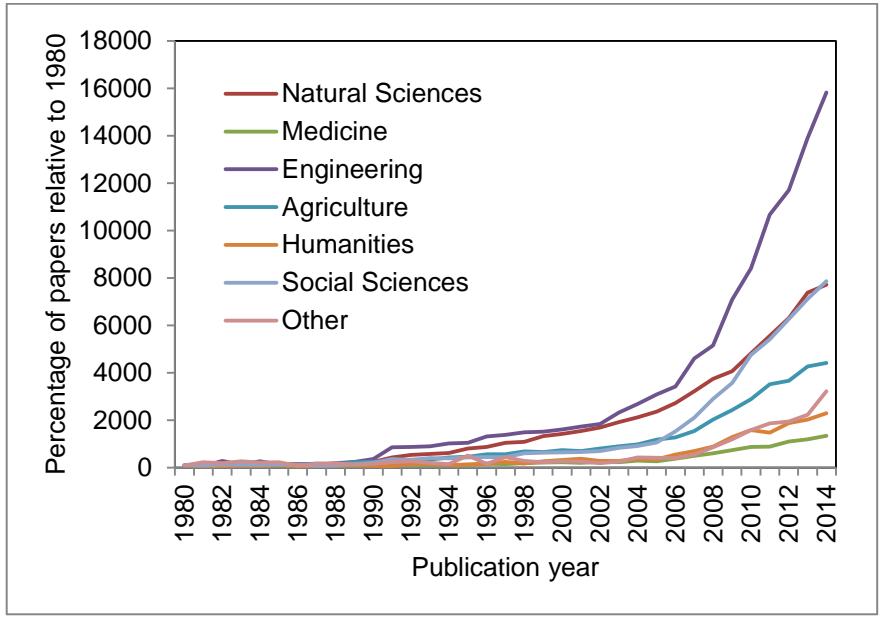

FIG 3: Field-specific relative increase of climate change related papers published since 1980, based on the main OECD categories (the number of papers published in 1980 equals 100%).

According to FIG 4, climate change research has become an issue also for disciplines beyond the natural sciences (e.g. engineering, history, law, management, sociology etc.). The categories *Engineering* and *Social Sciences* show the strongest increase since around 2005. Since around 2009 the relative increase of the *Natural Sciences* and the *Social Sciences* is almost identical. Obviously, climate change is increasingly seen as a fact to be considered for the near future: The need to limit fuel combustion and to adapt to global warming apparently is a huge stimulation for various technological developments and research on the implications of climate change. For example, sociologists analyzed the public understanding and the discussion of climate change in science, politics, and the mass media (Weingart et al., 2000; McCright and Dunlap 2011). Furthermore, the *Humanities* have discovered climate change as a research topic. Historians for example reconstructed climate extremes in medieval history (Pfister, 2007).

Climate change affects agriculture in a number of ways (changes in average temperatures, rainfall, climate extremes) and therefore has become an important field of investigation within climate change research. Climate change will likely affect food production and probably increases the risk of food shortage (IPCC Synthesis Report 2014, Summary for Policymakers, p. 15). Although climate change is increasingly relevant also for medicine, the portion of output of this field of study is comparatively low.

The OECD categories are very broad and imply a classification of journals and not of the specific papers published therein. Also, many journals publish papers from different research fields and are assigned to more than one category which causes a certain amount of impreciseness. In order to differentiate our publication set with regard to the major subfields of climate change research we applied a different method: We parsed our paper set by combining with carefully selected further search terms which are based (1) on the title word analysis from step 1 of our literature search procedure and (2) on the major topics of climate change research as indicated by various summarizing publications (e.g. the IPCC Synthesis Report 2014, table of contents).

As a first category of search terms, we selected the papers dealing with the main climate subsystems: the atmosphere, the oceanic water, the continental water, the ice sheets and glaciers, and the continental biomass. Additionally, we selected the literature specifically dealing with the various forms of atmospheric and oceanographic circulation or oscillation phenomena. All these search terms mark the kind of basic research in climatology, atmospheric- and geosciences, meteorology, and oceanography, which is undertaken to better understand the earth's climate system. We completed this category by separately searching for the more theoretical publications dealing with climate modeling and the prediction of future climate. As a second category, we searched for papers dealing with the adaptation to climate change or its mitigation as well as papers focusing on effects, impacts, and risks of climate change. Such research takes climate change more or less as a matter of fact and discusses possible consequences and reactions. The corresponding terms were searched in titles and keywords only, because a search in abstracts might have resulted in too many false positives. For more detailed information of the search procedure, see the search terms in Table 1. FIG 4 shows the time evolution of the papers of the major subfields within climate change research.

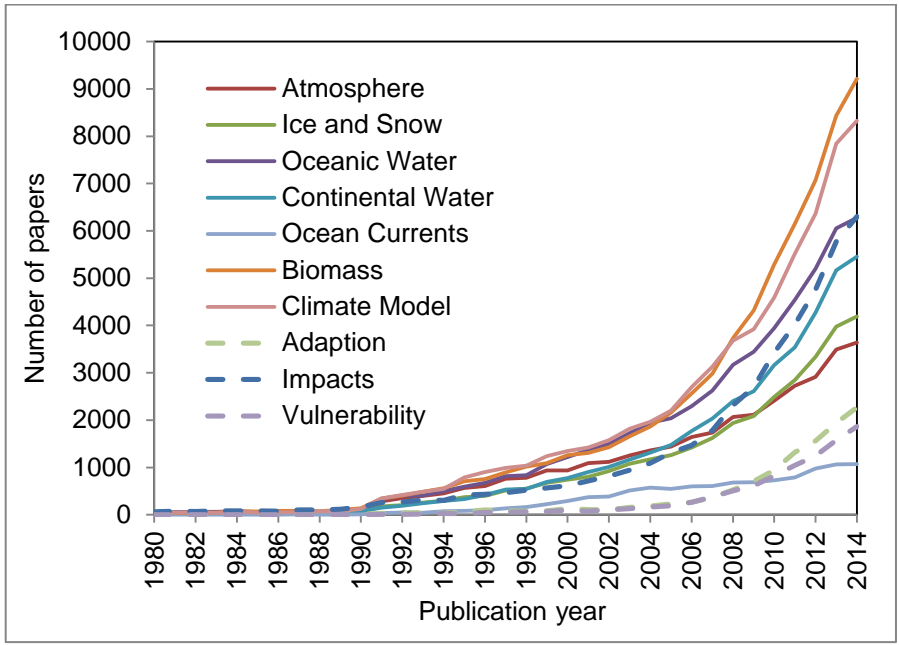

FIG 4: Time evolution of the papers of the major subfields within climate change research. These research topics comprise 182,594 papers (82.2%) of our publication set (n = 222,060).

According to FIG 4, continental biomass related research is the largest subfield within climate change research, closely followed by climate modeling, which demonstrates the importance of theoretical investigations (admittedly, these two subfields are rather broad). Next come research dealing with oceanic water, with impacts and effects of climate change, and with continental water (lakes, rivers, rainfall). Due to the radiative imbalance of the earth, less energy leaves the atmosphere than enters it. By far most of this extra energy has been absorbed by the oceans, which makes the oceans a major climate factor. The term "*sea*" was searched in addition to the terms "*ocean*" and "*marin*" to include papers dealing for example with changing sea surface temperatures or the rise of the sea levels into the answer set. Another major subfield is represented by the topic ice and snow (e.g. ice cores, ice sheets, glaciers, shelf ice). Ice cores are most important for the dating and reconstruction of the earth's past climate as well as for the prediction of the future climate.

The next major subfield is related to the atmosphere as another important climate subsystem (beside the ice and water related subsystems). This subfield includes research on clouds, on wind and storms (i.e. the key topics of meteorology), but also on aerosols (see volcanic eruptions). In contrast to impacts and effects of climate change, which appear as a major field of interest, research dealing with the adaptation to and the mitigation of climate change as well as

17with the risks and the vulnerability of global warming are comparatively small. Both were next to insignificant until 2004, but their share of papers increased exponentially since 2005, showing the strongly increasing research activity in this field. Global warming also affects ocean currents and thereby periodical climate changes like ENSO (El Nino Southern Oscillation) and NAO (North Atlantic Oscillation). As a more specific research topic and a subset of the atmospheric and oceanic water subfields, this research represents the smallest topic within our publication set, thereby masking somewhat the importance of these research activities. In Table 2, the total number of papers as well as the bibliometric indicators $PP_{Top\ 50\%}$, $PP_{Top\ 10\%}$, and $PP_{Top\ 1\%}$ of the papers belonging to the specific subfields are given.

Table 2: Major subfields of climate change research, ranked by publication output within the time period 1980-2012. In addition, the bibliometric indicators $PP_{Top\ 50\%}$, $PP_{Top\ 10\%}$, and $PP_{Top\ 1\%}$ of the papers belonging to the specific subfields are given.

| Climate change subfield | # Papers | $PP_{Top\ 50\%}$ | $PP_{Top\ 10\%}$ | $PP_{Top\ 1\%}$ |
|---|---|---|---|---|
| Biomass | 48,823 | 68.68 | 20.83 | 2.92 |
| Climate models | 47,623 | 67.43 | 19.96 | 2.74 |
| Oceanic water | 41,356 | 68.50 | 20.76 | 2.97 |
| Impacts | 30,503 | 65.22 | 18.39 | 2.38 |
| Continental water | 30,078 | 67.91 | 18.49 | 2.10 |
| Atmosphere | 28,337 | 68.25 | 20.53 | 3.00 |
| Ice and snow | 25,481 | 70.02 | 20.72 | 2.66 |
| Ocean currents | 8,686 | 69.74 | 21.70 | 2.89 |
| Adaptation | 7,254 | 66.78 | 21.57 | 3.47 |
| Vulnerability | 5,967 | 67.22 | 21.94 | 3.51 |
| All subfields | 137,586 | 66.11 | 19.00 | 2.55 |
| Other climate change literature | 31,967 | 51.89 | 12.36 | 1.52 |
| All climate change literature | 169,553 | 63.43 | 17.75 | 2.36 |

According to Table 2, all subfields together comprise more than 81% of the total climate change papers published within the time period 1980-2012 (note that we have restricted all citation impact analyses to 2012 as the most recent publication year). The $PP_{Top\ 50\%}$ values of all subfields are above the proportion of the total climate change literature ($PP_{Top\ 50\%}$=63.43%).

Research on vulnerability ($PP_{Top\ 1\%}$= 3.51) and on adaptation ($PP_{Top\ 1\%}$= 3.47) can be seen as the subfields within climate change research publishing the largest proportion of very important papers.

**3.2 Contributing journals**

In accordance with the publication practice in the core natural sciences, we assume that research results from climate change research are mainly published as journal (or conference) articles which are predominantly covered by literature databases like WoS. Thus, the number of papers published in a specific journal can be seen as a measure of the importance or "weight" of that journal for a specific research topic or field. In so far, it is interesting to find out, which journals are dominating quantitatively as publication medium for researchers active in the field of climate change research. Table 3 shows the distribution of the climate change research papers included in our data set throughout the journals which have published at least 1000 papers. Again, the bibliometric indicators $PP_{Top\ 50\%}$, $PP_{Top\ 10\%}$, and $PP_{Top\ 1\%}$ have been calculated.

Table 3: Distribution of climate change research papers throughout the top journals (with at least 1,000 papers). In addition, the bibliometric indicators $PP_{Top\ 50\%}$, $PP_{Top\ 10\%}$, and $PP_{Top\ 1\%}$ of the journals are given.

| Journal | # Papers | $PP_{Top\ 50\%}$ | $PP_{Top\ 10\%}$ | $PP_{Top\ 1\%}$ |
|---|---|---|---|---|
| Journal of Geophysical Research | 6,156 | 73.76 | 20.64 | 2.17 |
| Journal of Climate | 4,109 | 79.68 | 28.82 | 3.95 |
| Geophysical Research Letters | 3,754 | 79.34 | 27.45 | 4.02 |
| Climatic Change | 2,333 | 69.90 | 17.47 | 2.68 |
| Palaeogeography, Palaeoclimatology, Palaeoecology | 2,178 | 79.53 | 21.15 | 1.76 |
| Climate Dynamics | 1,835 | 75.46 | 23.41 | 2.29 |
| International Journal of Climatology | 1,777 | 65.09 | 13.44 | 1.80 |
| Global Change Biology | 1,722 | 93.17 | 43.71 | 5.83 |
| Quaternary Science | 1,644 | 87.97 | 33.95 | 4.53 |
| Reviews | | | | |
| Energy Policy | 1,403 | 63.83 | 13.67 | 1.02 |
| Nature | 1,064 | 98.47 | 37.80 | 3.75 |
| Quaternary Research | 1,051 | 80.43 | 23.47 | 1.18 |
| Global and Planetary Change | 1,043 | 69.09 | 19.20 | 2.17 |

Most important are the *Journal of Geophysical Research*, the *Journal of Climate*, and the *Geophysical Research Letters* in terms of the total number of papers published. The journal *Climatic Change*, which has been founded specifically for research papers on climate change, appears on rank four. *Nature* as one of the most prominent multidisciplinary journals appears on lower ranks but shows the highest $PP_{Top\,50\%}$: Nearly all papers published in *Nature* belong to the 50% most frequently cited papers. Most journals show a comparatively high citation impact. The proportion of highly received papers ($PP_{Top\,1\%}$) is very large for the journals *Global Change Biology* ($PP_{Top\,1\%}$=5.83%) and *Quaternary Science Reviews* ($PP_{Top\,1\%}$=4.53%).

## 3.3 Contributing countries

Climate change research is not only a highly multidisciplinary undertaking but also a research area with many countries being active and cooperating with each other. The number of papers of each country and their citation impact based on the $PP_{Top\,50\%}$ values are shown in Table 4, together with the percentage of excellent papers (i.e. $PP_{Top\,10\%}$ and $PP_{Top\,1\%}$). The $PP_{TopX\%}$ values in columns 2 are relative to the countries' overall impact of all papers between 1980 and 2012. A value of 200 for example corresponds to twice the impact of the countries' climate change papers compared to all the countries' papers in the aforementioned time frame.

Table 4: Countries of authors ranked by publication output within the time period 1980-2012 (only countries with at least 1,000 papers are considered). All contributing authors are considered; this implies a substantial overlap, since the cooperating authors of a specific paper often work in different countries. In addition, the bibliometric indicators $PP_{Top\,50\%}$, $PP_{Top\,10\%}$, and $PP_{Top\,1\%}$ of the countries are included: Column (1) includes the impact of climate change papers, and column (2) displays the impact of climate change papers relative to the overall impact of the countries' papers published between 1980 and 2012.

| Country | # Papers | PP$_{Top\ 50\%}$ | | PP$_{Top\ 10\%}$ | | PP$_{Top\ 1\%}$ | |
|---|---|---|---|---|---|---|---|
| | | (1) | (2) | (1) | (2) | (1) | (2) |
| USA | 61,941 | 72.06 | 179.51 | 23.67 | 237.17 | 3.66 | 316.82 |
| UK | 21,777 | 74.59 | 190.84 | 26.13 | 294.99 | 4.13 | 423.78 |
| Germany | 14,971 | 69.80 | 172.02 | 22.61 | 265.55 | 3.44 | 390.66 |
| Canada | 13,499 | 69.00 | 162.22 | 20.46 | 222.56 | 3.19 | 320.96 |
| China | 11,185 | 56.98 | 138.46 | 14.58 | 192.44 | 1.60 | 232.20 |
| France | 10,999 | 69.82 | 167.18 | 23.04 | 269.76 | 3.72 | 430.53 |
| Australia | 10,555 | 70.95 | 160.65 | 23.53 | 252.41 | 4.05 | 392.68 |
| Spain | 6,395 | 67.46 | 161.26 | 19.05 | 245.30 | 2.84 | 378.67 |
| Japan | 6,172 | 56.50 | 145.66 | 14.28 | 227.78 | 2.07 | 392.35 |
| Netherlands | 6,116 | 75.16 | 156.01 | 26.85 | 232.08 | 4.80 | 359.68 |
| Italy | 5,580 | 67.90 | 165.49 | 20.96 | 264.34 | 3.31 | 413.01 |
| Switzerland | 5,269 | 78.38 | 162.76 | 29.92 | 240.29 | 5.00 | 325.08 |
| Sweden | 5,079 | 74.19 | 149.48 | 25.06 | 229.48 | 3.87 | 329.09 |
| Norway | 3,964 | 74.07 | 154.95 | 24.66 | 247.47 | 3.92 | 358.30 |
| India | 3,807 | 42.46 | 137.73 | 9.09 | 222.20 | 0.97 | 292.11 |
| Russia | 3,340 | 41.65 | 185.92 | 11.56 | 373.52 | 1.66 | 565.97 |
| Denmark | 3,060 | 76.59 | 154.37 | 27.36 | 232.72 | 5.14 | 369.13 |
| Brazil | 2,991 | 52.99 | 160.87 | 14.96 | 325.61 | 2.13 | 551.35 |
| Finland | 2,743 | 71.93 | 148.36 | 21.64 | 220.25 | 3.40 | 326.26 |
| New Zealand | 2,427 | 71.12 | 163.09 | 20.99 | 248.24 | 3.39 | 352.48 |
| Belgium | 2,398 | 71.34 | 160.87 | 21.51 | 217.19 | 3.56 | 324.99 |
| South Africa | 2,161 | 61.91 | 170.51 | 17.66 | 307.81 | 3.17 | 551.06 |
| Austria | 2,021 | 69.30 | 171.49 | 23.02 | 271.01 | 4.66 | 478.12 |
| South Korea | 1,764 | 53.64 | 137.88 | 10.70 | 171.91 | 1.71 | 329.94 |
| Israel | 1,710 | 64.22 | 145.33 | 16.43 | 185.69 | 2.55 | 283.49 |
| Argentina | 1,562 | 59.71 | 173.14 | 13.84 | 285.70 | 1.79 | 423.92 |
| Mexico | 1,367 | 52.21 | 155.56 | 13.61 | 275.81 | 2.04 | 435.07 |
| Poland | 1,324 | 51.68 | 161.18 | 12.90 | 289.66 | 1.94 | 443.48 |
| Greece | 1,309 | 61.02 | 158.80 | 16.14 | 243.76 | 2.53 | 406.74 |
| Portugal | 1,308 | 70.09 | 158.63 | 22.07 | 266.46 | 4.00 | 485.04 |
| Taiwan | 1,272 | 56.30 | 130.39 | 13.33 | 201.02 | 1.09 | 228.89 |
| Turkey | 1,192 | 47.93 | 155.34 | 11.78 | 271.81 | 2.21 | 607.25 |

According to Table 4, research on climate change is quantitatively dominated by the USA, followed by the UK, Germany, and Canada. China appears on rank five, followed by France and Australia. $PP_{Top\ 50\%}$ of these seven countries (with more than 10,000 papers in total) extends between 56.98% (China) and 74.59% (UK). $PP_{Top\ 1\%}$ ranges from 1.6% (China) to 4.13% (UK). $PP_{Top\ 10\%}$ ranges from 14.58% (China) to 26.13% (UK). Hence, the three citation-based indicators exhibit consistently that the UK has produced papers in climate change research with the largest reception compared to the other countries (with more than 10,000 papers). However, the other top countries rank nearby (with the exception of China, which nevertheless ranks above average). Switzerland, Denmark and also The Netherlands (with a publication output between around 3,000 and 6,000 papers) perform top with regard to all three bibliometric indicators – the impact of their contributions to climate change research is impressive. The citation impact of the climate change papers of all countries is above or far above the overall impact of the countries' papers each.

Li et al. (2011) presented a comparison of publication trends of the top seven most productive countries and found a quite similar ranking concerning the publication numbers (with only one exception: Australia appears on rank 5 compared to rank 7 in our publication output ranking).

**3.4 Visualization of the time evolution of research topics**

The maps presented in FIG 5 and FIG 6 show the title word clusters (clouds) of the climate change papers of the overall publication set (1980-2014) and of the papers from three specific publication time periods (1980-1990, 2003, and 2014). We have chosen these specific time periods in order to compare the papers of the most recent (complete) publication year (2014) with early publication years (1980-1990) and a publication year in between (2003). Due to the low number of papers per year before 1990 (caused by both the low publication output at that time and the lack of abstracts in WoS prior to 1991) we had to accumulate the early papers from a publication time period of about a decade (1980-1990). All title words of the same cluster appear as circles with the same color. The distance between the circles relates to the distance (or closeness) in terms of bibliographic coupling. The size of the circle is proportional to the number of papers found with these terms in the titles.

FIG 5: Title words from bibliographic coupling of climate change papers published within the overall time period 1980-2014. The minimum number of papers containing a specific title word is 20. Readers interested in an in-depth analysis of our publication set can use VOSviewer interactively and zoom into the clusters. We gladly provide the necessary NET and MAP files for VOSviewer at Haunschild, Bornmann, & Marx (2015a).

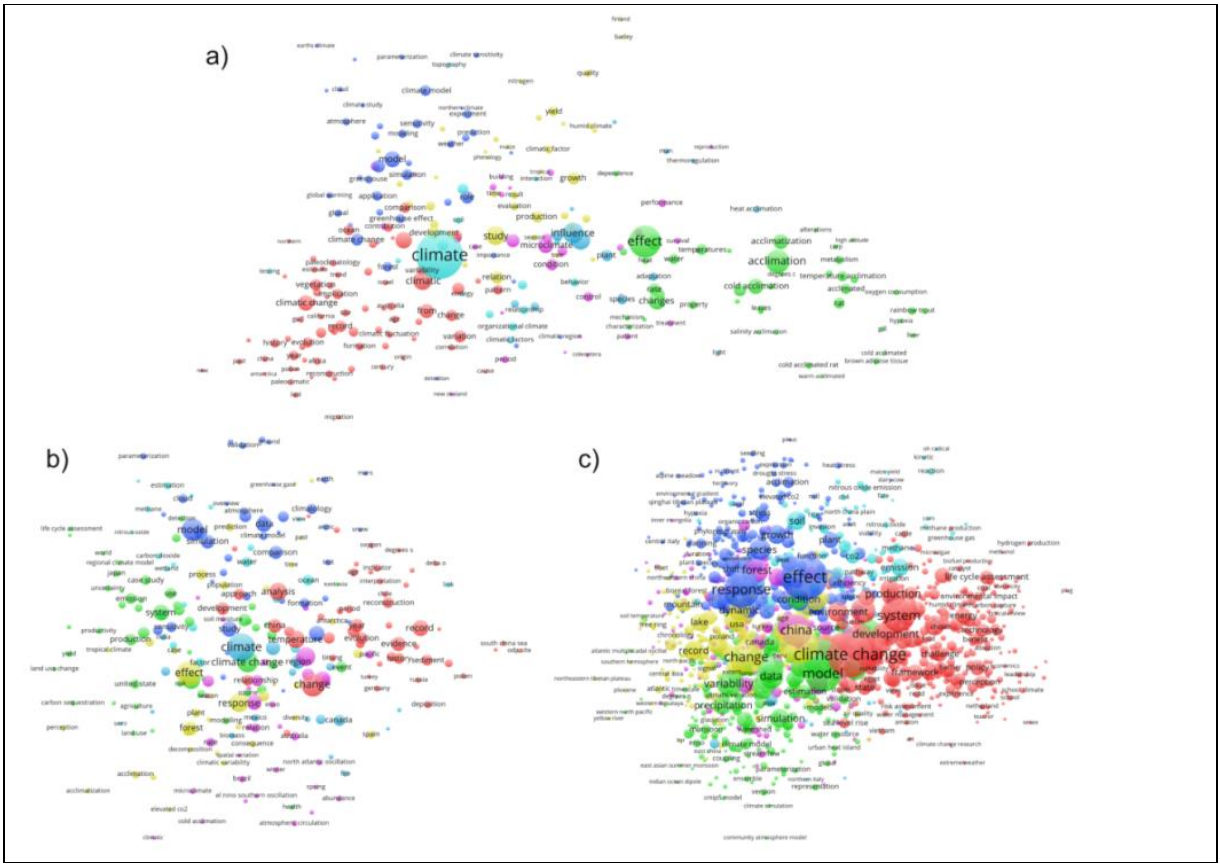

FIG 6: Title words from bibliographic coupling of climate change papers published a) 1980-1990, b) 2003, and c) 2014. The minimum number of papers containing a specific title word is 10. The coloring of the clusters automatically performed by VOSviewer is different for each map of the three selected publication times. Readers interested in an in-depth analysis of our publication set can use VOSviewer interactively and zoom into the clusters. We gladly provide the necessary NET and MAP files for VOSviewer at Haunschild, Bornmann, & Marx (2015b).

The most pronounced title words of the papers published within the overall time period 1980-2014 are *climate change*, *effect*, and *impact* (center). The red cluster (center right) includes papers related to energy and policy topics. Major title words are: *climate change*, *adaptation*, *emission*, *framework*, *uncertainty*, *cost*, *technology*, and *policy* (in the order of decreasing frequency). The blue cluster (center left) combines the papers around paleoclimate. Major title words are: *record*, *year*, *variation*, *lake*, *sediment*, and *event*. The green cluster (bottom left) contains theoretical publications. Major title words are: (*climate*) *model*, *data*, *parametrization*, and *simulation* (but also *variability*, which often appears in combination with climate modeling). The yellow cluster (center top) illustrates the importance of biological effects of global warming.

Major title words are: *effect*, *forest*, *soil*, and *plant*. And finally, the magenta cluster (top) marks papers concerning acclimatization and survival of species.

If we analyze and compare the maps based on the three selected time spans (c.f. FIG 6) we find some remarkable changes: The title word map of the first decade (1980-1990, FIG 6a) shows the term *climate* as the most pronounced title word. The terms *effect* and *influence* appear secondarily. The term *climatic change* and the related terms appear third-rated (i.e. as small circles). The title word map constructed from papers published in the year 2003 (FIG 6b) for the first time accentuates the term *change*. The 2014 map (FIG 6c) is quite similar to the map of the overall publication set (FIG 5) with *climate change* and *effect* as the most pronounced terms. The reader might miss the term *impact*, but it is hidden behind *climate change*.

The changing title words based on the maps of the three specific publication times exhibit that the term *climate change* comes forward with time. Obviously, the authors increasingly use a term which implies global warming (and therewith anthropogenic causes) as a matter of fact. Furthermore, the term *impact* arises and points to research dealing with the various effects and risks of climate change – see also the IPCC Synthesis Report 2014, Summary for Policymakers. The term *model* and related terms (e.g. *simulation*) appear independently of time. This indicates the high relevance of climate modeling since the beginning of the time period analyzed here.

## 4. Discussion

This bibliometric study of a large and carefully selected publication set of papers dealing with research on climate change presents a multitude of quantitative data: (1) The growth of the overall publication output of climate change research as well as (2) of some major subfields, (3) the contributing journals and countries and their citation impact, and (4) a title word analysis aiming to illustrate the time evolution and relative importance of specific research topics.

The total number of papers dealing with climate change shows a strong increase: Within the time period 1991 to 2010, the number of climate change papers increased by a factor of ten and exhibits a doubling every 5-6 years. The exponential growth of climate change literature is possibly induced by the increasing influence of the IPCC Assessment Reports, which underlined risks of global warming and eventually made climate change research a hot topic. These reports revealed the strong need of further research for a better understanding of the earth's climate system and for improved predictions of the future climate. Our findings are in rough accordance with Grieneisen and Zhang (2011), who reported that the number of publications on climate change and global warming has doubled with a rate of approximately every 4 years. In contrast, Stanhill (2001) had found a doubling rate of 11 years. But his publication set is based only on the abstract journal of the American Meteorological Society from the (earlier) time period 1951-1997. Compared with the growth of the overall science, the growth rate of climate change related publications is extraordinarily high: The total volume of publications covered by the WoS between 1980 and 2012 doubled approximately only every 24 years (Bornmann & Mutz, 2015).

According to our subfield analysis, continental biomass related research is the major subfield within climate change research, closely followed by climate modeling. Next come research dealing with oceanic water, with impacts and effects of climate change, and with continental water (lakes, rivers, rainfall). Another major subfield is represented by the topics ice and snow (ice cores are most important for the dating and reconstruction of the earth's past climate). The next major subfield is related to the atmosphere as another important climate subsystem (including research on clouds, wind, and storms). Research dealing with adaptation, mitigation, risks, and vulnerability of global warming is comparatively small, but their share of papers increased exponentially since 2005. As a more specific research topic and a subset of the oceanic water subfield, research on ocean currents represents the smallest topic within our publication set. The normalized citation impact of all subfields measured in terms of the proportion of most frequently cited papers is significantly above the expected values (50%, 10%, and 1%) and also above the proportions of the total climate change literature. Research on vulnerability ($PP_{Top\ 1\%}$= 3.51) and on adaptation ($PP_{Top\ 1\%}$= 3.47) published the largest proportion of very important papers for climate change research.

The journal analysis of our publication set revealed that the *Journal of Geophysical Research*, the *Journal of Climate*, and the *Geophysical Research Letters* appear at the top positions of the publication output ranking (in this order). The journal *Climatic Change*, which has been founded specifically for research papers on climate change, appears on rank four. *Nature* as one of the most prominent multidisciplinary journals appears on a lower rank but shows a very high citation impact.

Research on climate change is quantitatively dominated by the USA followed by the UK, Germany, and Canada. China appears on rank five, followed by France and Australia. The $PP_{Top\ 50\%}$ values of these seven countries (with more than 10,000 papers in total) are between 56.98 (China) and 74.59 (UK). The three citation-based indicators exhibit consistently that the UK has produced papers in climate change research with the largest reception compared to the other countries (with more than 10,000 papers). Also, Switzerland, Denmark, and The Netherlands (with a publication output between around 3,000 and 6,000 papers) perform on a high level with regard to the three bibliometric indicators.

We mention here that the literature output can be seen as a combined measure of the size of a specific subfield as well as the amount of research activity and that it is no measure of research performance. For example, research dealing with ice is sometimes connected with highly specific research methods like ice core dating, which can be executed by only a few drilling teams. Although most important for the reconstruction of the earth's past climate, the publication volume of such research is comparatively low.

The title word analysis shows that the term *climate change* comes forward with time. Obviously, the authors increasingly use a term which implies global warming (and therewith anthropogenic causes) as a matter of fact. Furthermore, the term *impact* arises and points to research dealing with the various effects of climate change. The discussion of the question of human induced climate change towards a clear fact (for the majority of the scientific community) stimulated research on future pathways for adaptation and mitigation. Finally, the term *model* and related terms (e.g. *simulation*) appear independently of time, indicating the high relevance of climate modeling also revealed by the subfield analysis.

This study is a first attempt to a mapping of the complete climate change literature. However, more bibliometric research is needed to analyze and overview the research field from a quantitative perspective. Future research should focus on in-depth analyses of more specific topics like the impact of global warming on agriculture, fishery, forestry, and viniculture (winegrowing). Such studies can contribute to an understanding of the evolution, structure, and knowledge base of climate change research.

Like most bibliometric analyses this study has some limitations to be mentioned here: (1) The completeness of our data set is limited by the fact that abstracts are not searchable in WoS prior to 1991. (2) Title words are sometimes multi-meaning and not sufficiently specific for detailed interpretations. Therefore, one should avoid over-interpretation of title word analyses via VOSviewer.

**Acknowledgements**

The bibliometric data used in this paper are from an in-house database developed and maintained by the Max Planck Digital Library (MPDL, Munich) and derived from the Science Citation Index Expanded (SCI-E), Social Sciences Citation Index (SSCI), Arts and Humanities Citation Index (AHCI) prepared by Thomson Reuters (Philadelphia, Pennsylvania, USA).

30Li, J., Wang, M.H. and Ho, Y.S. (2011). Trends in research on global climate change: A science citation index expanded-based analysis. Global and Planetary Change 77, 13-20. http://dx.doi.org/doi:10.1016/j.gloplacha.2011.02.005

Li, W. and Zhao, Y. (2015). Bibliometric analysis of global environmental assessment research in a 20-year period. Environmental Impact Assessment Review 50, 158–166. http://dx.doi.org/doi:10.1016/j.eiar.2014.09.012

Marx, W. and Bornmann, L. (2015). On the causes of subject-specific citation rates in Web of Science. Scientometrics 102, 1823-1827. http://dx.doi.org/doi:10.1007/s11192-014-1499-9

McCain, K.W. (2012a). Assessing obliteration by incorporation: Issues and caveats. Journal of the Association for Information Science and Technology 63(11), 2129-2139. http://dx.doi.org/doi:10.1002/asi.22719

McCain, K.W. (2012b). "Obliteration by Incorporation" in B. Cronin & C. R. Sugimoto (Eds.), *Beyond Bibliometrics harnessing multi-dimensional indicators of performance*. Cambridge, MA, USA: MIT Press, p. 129-149.

McCright, A.M. and Dunlap, R.E. (2011). The politicization of climate change and polarization in the American public's views of global warming, 2001-2010. Sociological Quarterly 52(2), 155-194. http://dx.doi.org/doi:10.1111/j.1533-8525.2011.01198.x

Merton, R.K. (1965). On the shoulders of giants: A shandean postscript. New York, The Free Press.

Pasgaard, M. and Strange, N. (2013). A quantitative analysis of the causes of the global climate change research distribution. Global Environmental Change 23, 1684-1693. http://dx.doi.org/doi:10.1016/j.gloenvcha.2013.08.013

Pfister, C. (2007). Climatic extremes, recurrent crises and witch hunts: Strategies of European societies in coping with exogenous shocks in the late sixteenth and early seventeenth centuries. Medieval History Journal 10(1-2), 33-73. http://dx.doi.org/doi:10.1177/097194580701000203

Schwechheimer, A. and Winterhager, M. (1999). Highly dynamic specialities in climate research. Scientometrics 44(3), 547-560. http://dx.doi.org/doi:10.1007/BF02458495